\documentclass[12pt]{article}
\usepackage[utf8]{inputenc}
\usepackage{amsfonts}
\usepackage{amsmath}
\usepackage{amsthm}
\usepackage{verbatim}
\usepackage{nameref}
\usepackage[colorlinks,citecolor=blue,urlcolor=blue,linkcolor=blue]{hyperref}
\usepackage{natbib}
\usepackage{graphicx}
\usepackage{float}

\usepackage{algorithm}
\usepackage{algorithmic}
\usepackage[algo2e]{algorithm2e}
\restylefloat{algorithm}

\title{Efficiently resolving rotational ambiguity in Bayesian matrix sampling with matching}
\author{Evan Poworoznek$^{*}$ \and Niccolo Anceschi$^{*}$ \and Federico Ferrari$^{*,\dag}$ \and David Dunson$^{*}$}
\date{\textit{$^{*}$Duke University, Department of Statistical Science\\%
    $^{\dag}$Harvard T.H. Chan School of Public Health, Harvard}}
\begin{document}

\maketitle

\begin{abstract}
A wide class of Bayesian models involve unidentifiable random matrices that display rotational ambiguity, with the Gaussian factor model being a typical example. A rich variety of Markov chain Monte Carlo (MCMC) algorithms have been proposed for sampling the parameters of these models. However, without identifiability constraints, reliable posterior summaries of the parameters cannot be obtained directly from the MCMC output. As an alternative, we propose a computationally efficient post-processing algorithm that allows inference on non-identifiable parameters. We first orthogonalize the posterior samples using Varimax and then tackle label and sign switching with a greedy matching algorithm. We compare the performance and computational complexity with other methods using a simulation study and chemical exposures data. The algorithm implementation is available in the \texttt{infinitefactor} \textbf{\textsf{R}} package on CRAN. 
\end{abstract}
\noindent%

\section{Introduction}\label{intro}
%A wide class of Bayesian models for matrices involve non-identifiability in the parameters, with a typical example being the Gaussian factor model, refer, for example, to \citep{Mike03}:
Factor models are commonly used to characterize the dependence structure in correlated variables while providing dimensionality reduction. With continuous observations, this is usually achieved by assuming that the observed variables are linear combinations of a set of lower dimensional latent variables. Let $X_i = (x_{i1},\ldots, x_{ip})^T$ be a $p \times 1$ vector of correlated variables. Without loss of generality, we assume that the data have been centered prior to analysis so that we can omit the intercept. The typical Gaussian factor model has the following representation: 
\begin{gather} \label{fact_model}
\begin{split}
    X_i & = \Lambda \eta_i + \epsilon_i, \quad \epsilon_i \sim N(0,\Sigma), \\
    \eta_i & \sim  N_k (0,I),
\end{split}
\end{gather}
where $\Lambda$ is a $p \times k$ tall and skinny matrix of factor loadings, $\eta_i$ is a $k \times 1$ vector of latent factors and $\Sigma = \text{diag}(\sigma_1^2, \ldots, \sigma_p^2)$ is a diagonal matrix containing the residual variances. A typical choice for the latent dimension is to set $k \ll p$ in order to provide dimensionality reduction. The factor model specification induces a decomposition of the $p \times p$ covariance matrix of $X_i$:
\begin{align}\label{cov_decomposition}
    \text{cov}(X_i) = \Lambda\Lambda^T + \Sigma.
\end{align}
This decomposition implicitly assumes that the correlation among variables is fully explained by the latent variables via the term $\Lambda \Lambda^T$.

It is well known that the covariance decomposition in $(\ref{cov_decomposition})$ is not unique, and as a result $\Lambda$ is non-identifiable. For example, consider a $k \times k$ semi-orthogonal matrix $P$ ($P P^T = I$), then $\Lambda^{'} = \Lambda P$ also satisfies the above equation. We refer to this as rotational ambiguity throughout the paper. %(\textbf{mention sign switching} before). 
%A first approach to ident of factor models was first offered by \cite{anderson1956statistical}. 
Although the estimation of some parameters, such as the covariance, is not affected by this \citep{Bhattacharya2013}, rotational ambiguity is still of crucial importance as it prevents inference on $\Lambda$ and interpretation of the induced groupings of variables from (\ref{fact_model}).

Non-identifiability of the factor loadings matrix is a well studied problem in the frequentist literature, where non-identifiability causes multimodality in the objective function. However, from a statistical perspective, each mode is equivalently optimal \citep{lawley1962factor} and it suffices to choose a single mode to perform inference. A possibility to select one mode is by applying an orthogonalization procedure, such as Varimax \citep{varimax}, Quartimax \citep{quartimax}, Promin \citep{promin}, or the Orthogonal Procrustes algorithm in \cite{BADFM}. %These methods are well studied and commonly used in factor analysis in frequentist . 
However, in the Bayesian paradigm, multimodality in the posterior distribution of the factor loadings matrix allows the sampling of $\Lambda$ to switch between modes during the MCMC procedure, effectively preventing convergence to a single mode. 

One possibility to attain identifiability of $\Lambda$ in Bayesian models is by imposing significant constraints \citep{Ghosh}, \citep{erosheva2011dealing}. A typical choice to enforce identifiability requires setting the upper diagonal elements of $\Lambda$ equal to zero and requiring the diagonal elements to be positive \citep{geweke1996measuring}. A more general solution was provided by \cite{fruhwirth2018sparse} using generalized lower triangular (GLT) matrices. This structure has been used routinely in numerous settings; see, for example, \cite{lucas2006sparse}, in order to perform inference on $\Lambda$, at the cost of introducing order dependence among the factors \citep{carvalho2008high}. \cite{chen2020structured} show how structural information on the matrix of factor loadings affects identifiability and estimation of $\Lambda$. \cite{millsap2001trivial} show that fixing different elements of $\Lambda$ to zero in different runs of the algorithm can lead to convergence failure, or to a factor solution that is no longer in the same equivalence class. %in the case in which some observed variables load on multiple latent factors
 \cite{erosheva2011dealing} show that requiring loadings to be positive may result in nontrivial multimodality of the likelihood, and as a result chains with different starting values may produce solutions that are substantially different in fit. Finally, this structure on $\Lambda$ also constrains the class of estimable covariance matrices by forcing some entries of the matrix $\Lambda$ to be equal to zero.

To address rotational ambiguity, we propose to use a post-processing algorithm that allows for identification and interpretation of $\Lambda$. Crucially, this algorithm does not affect the choice of priors and structure for the matrix of factor loadings, which is a modeling choice left to the practitioner. Several post-processing algorithms have been proposed to deal with rotational ambiguity in specific settings. \cite{mcalinn2018dynamic} adapts an optimization procedure for posterior mode finding for Bayesian dynamic factor analysis in macroeconomic applications. \cite{kaufmann2017identifying} propose a post-processing clustering procedure in order to obtain an identified posterior sample and \cite{kaufmann2019bayesian} use an empirical procedure in order to identify a posterior mode by exploiting correlation among factors.
\cite{erosheva2011dealing} address sign switching across the samples of $\Lambda$ with an approach based on \cite{stephens2000dealing}. 

We address the post-processing task with a more general approach related to \cite{papastamoulis2020identifiability}  and \cite{PRA}. We divide the algorithm into an orthogonalization step and a sign permutation step. We first solve orthogonal ambiguity by performing Varimax \citep{varimax}, as in \cite{papastamoulis2020identifiability}, though our algorithm can be adapted to any orthogonalization procedure. The rotational ambiguity between samples of $\Lambda$ is then limited to switching in the column labels and column signs, which we will refer to as label switching and sign switching, respectively. We propose to solve both these problems by matching each posterior sample to a reference matrix, and using the matches to align the samples. The aligned samples can then be directly used for inference. By simplifying the alignment step and developing a greedy maximization procedure, we significantly improve the computational efficiency with respect to \cite{papastamoulis2020identifiability} and \cite{PRA}, while maintaining good estimation performance. Also, our method is not constrained to factor models with a latent dimension less than $50$, but can be used with a latent dimension in the order of hundreds. We focus our analysis on the latent factor model, but our algorithm can be applied to a much larger class of models involving rotational ambiguity in matrix valued parameters. 

We define the identifiability setting in Section \ref{section::ident_setting} and the MatchAlign algorithm in Section \ref{section::algorithm}. % We describe how to choose a reference matrix in Section \ref{sec::pivot} and provide the details of the matching procedure in Section \ref{sec::matching}.
We provide extensive simulations in Section \ref{sec::simulation} and analyze the performance of our post-processing algorithm compared to that of several alternative methods. In Section \ref{sec::genomic}, we apply our algorithm to genomics. Our algorithm is implemented in the \texttt{infinitefactor} \textbf{\textsf{R}} package on CRAN.  %\textbf{theory ? }

\section{Identifiability Setting} \label{section::ident_setting}

Let us consider the Gaussian Bayesian factor model introduced in $(\ref{fact_model})$.
%:
%\begin{gather*} \label{fact_model_2}
%\begin{split}
%    X_i & = \Lambda \eta_i + \epsilon_i, \quad \epsilon_i \sim N(0,\Sigma), \\
%    \eta_i & \sim  N (0,1),
%\end{split}
%\end{gather*}
%where $X_i = (x_{i1},\ldots, x_{ip})^T$ is a $p \times 1$ vector of correlated variables, $\Lambda$ is a $p \times k$ tall and skinny matrix of factor loadings, $\eta_i$ is a $k \times 1$ vector of latent factors and $\Sigma = \text{diag}(\sigma_1^2, \ldots, \sigma_p^2)$ is a diagonal matrix containing the residual variances. 
This model allows for dimensionality reduction in chararacterizing the covariance of $X_i$ through the decomposition
\begin{align*}
    \text{cov}(X_i) = \Lambda \Lambda^T + \Sigma.
\end{align*}
In order to tackle identifiability of $\Lambda$, we first need identification between $\Lambda \Lambda^T$ and the residual variance $\Sigma$. The identification of the residual variance is often overlooked in the factor model literature. A one-factor model is identifiable only if at least $3$ factor loadings are nonzero \citep{anderson1956statistical}, \citep{fruhwirth2018sparse}. For the remainder of the paper we will assume that $k \le \frac{p - 1}{2}$, which ensures that $\Sigma$ is identifiable when all rows of $\Lambda$ are nonzero. As a result, this condition guarantees identification of $\Lambda \Lambda^T$. See \cite{fruhwirth2018sparse} and \cite{papastamoulis2020identifiability} for a comprehensive analysis of the topic and for detailed conditions to ensure identification of $\Sigma$.

We now focus on uniqueness of the factor loading matrix $\Lambda$. As noted in the \nameref{intro}, the covariance decomposition is not unique, and as a result $\Lambda$ is non-identifiable. Consider a $k \times k$ semi-orthogonal matrix $P$ ($P P^T = I$), then $\Lambda^{'} = \Lambda P$ also satisfies \eqref{cov_decomposition}. We refer to this as rotational ambiguity. Rotational ambiguity is a well studied problem in the frequentist literature and can be solved with orthogonalization procedures such as Varimax \citep{varimax}.
In the Bayesian paradigm, there will typically be rotational drift as posterior samples are collected via an MCMC algorithms.  Such changes in rotation need to be adjusted for before calculating posterior summaries of $\Lambda$; this can be accomplished through a post processing approach to rotationally align the samples or via imposing restrictions on $\Lambda$ {\em a priori} to avoid the rotational drift entirely.

Under the latter approach, a typical choice is to restrict $\Lambda$ so that the upper triangular elements are zero and the diagonal elements are positive \citep{geweke1996measuring}. This restriction has been used routinely in numerous settings; see, for example, \cite{lucas2006sparse}, in order to perform inference on $\Lambda$, at the cost of introducing order dependence among the factors \citep{carvalho2008high}. A more general solution was provided by \cite{fruhwirth2018sparse} using generalized lower triangular (GLT) matrices. To address rotational ambiguity, we instead follow the former approach and propose a post-processing algorithm.

\section{Algorithm}\label{section::algorithm}

\begin{figure}[h]
    \centering
    \includegraphics[width=1\linewidth]{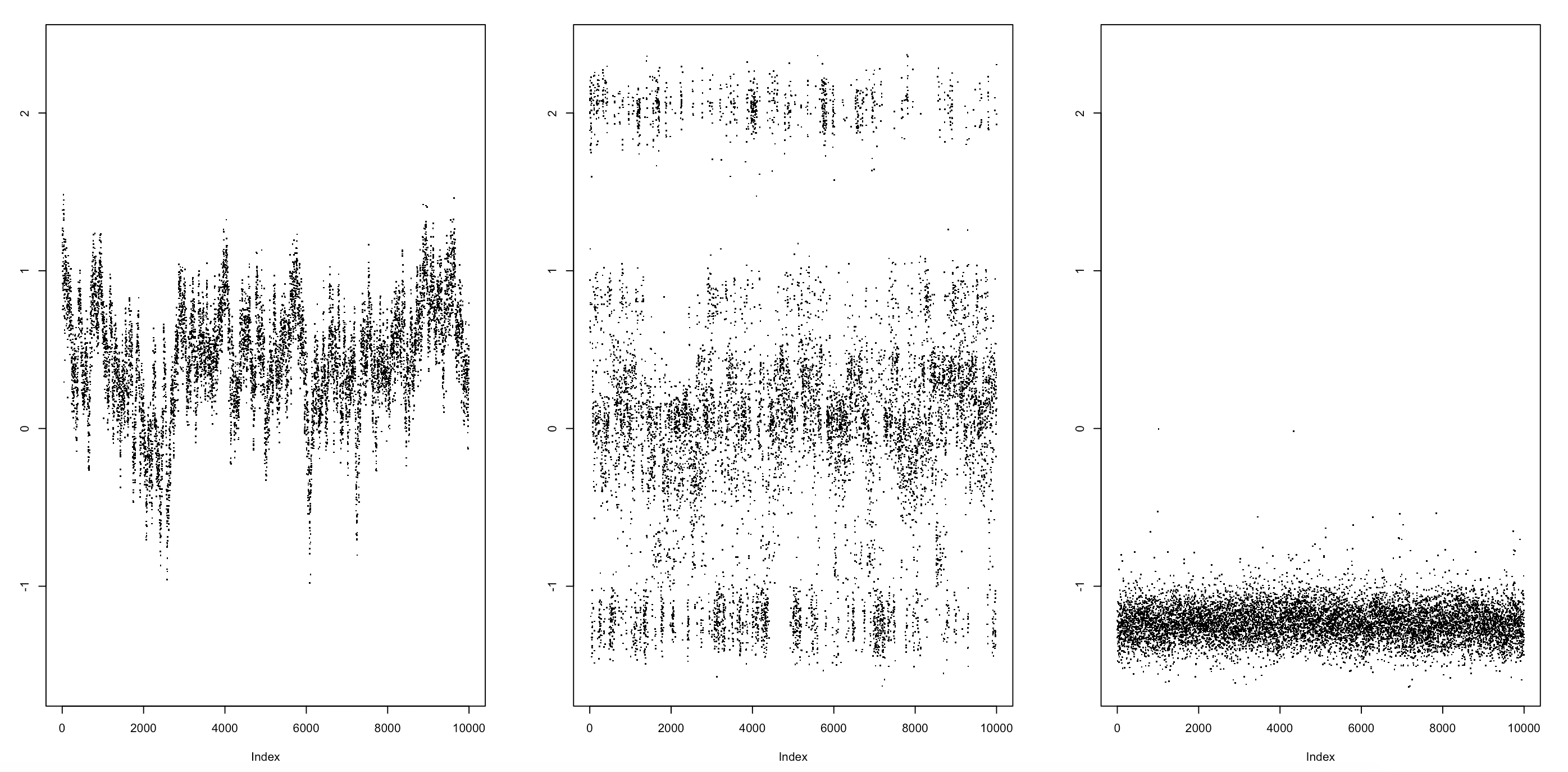}
    \caption{Symptoms of non-identifiability shown in trace plots of one $\Lambda$ entry showing full rotational ambiguity (left), sign and label switching after the application of Varimax (center), and complete alignment after applying MatchAlign (right)}
    \label{fig:trace_plots}
\end{figure}

In this section we describe our MatchAlign algorithm that solves rotational ambiguity in the posterior samples of non identifiable matrix value parameters. The MatchAlign procedure is summarized in Algorithm \hyperref[MA:Algo]{1}. The three key steps are: 1) apply an orthogonalization procedure to each posterior sample of $\Lambda$, 2) choose a reference matrix (the pivot) and 3) match the columns of each posterior sample to the pivot's columns. Figure \ref{fig:trace_plots} shows an example of the output of our algorithm. 

Notably the computation in the algorithm is split into two \texttt{for} loops that can each be massively parallelized or distributed. The orthogonalization step can be completed for each sample completely independently with no shared memory access in a distributed environment. This process is currently implemented in the \texttt{infinitefactor} with optional parallel computing using the \texttt{parallel} package in \textbf{\textsf{R}}. % and the efficient forking found in Mac and Linux based systems. 
The \texttt{for} loop to align the samples can be easily parallelized by assigning groups of samples to several machines, as long as each process has access to the pivot. Next, we describe the three main steps in the algorithm. 

\begin{algorithm}[p]
\caption{MatchAlign algorithm to solve rotational ambiguity in matrix valued parameters.} \vspace{0.5cm}
 \label{MA:Algo}
\SetAlgoLined
%\KwResult{Resolve rotational ambiguity in samples of a matrix valued random variable}
 \textbf{Input:} \{$\Lambda^{(t)}: t=1,...,T\}$\; \\
 1. \For{$t$ in $1:T$}{
 Orthogonalize $\Lambda^{(t)}$ using Varimax and output $\Tilde{\Lambda}^{(t)}$\;}
 2. Choose a pivot $\Lambda^P$ from \{$\Tilde{\Lambda}^{(t)}: t=1,...,T\}$ \;\\
 %3. Assign $p$ the pivot matrix and its negative catenated columnwise\;
 3. \For{$t$ in $1:T$}{
 \For{$j$ in $1:k$}{
 Compute normed differences between  $c_j^{(t)}$ and $\Lambda^P$ and $-\Lambda^P$ columns\; \\
 Retain the $j^{th}$ column having minimum norm value \\
 Drop the matched column and its negative from the pivot\;}
 Reorder and re-sign %$s$ according to $l$
 }
\end{algorithm}

%\subsection{Orthogonalization} \label{sub::orthogonalization}

\begin{figure}[h]
    \centering
    \includegraphics[width=1\linewidth]{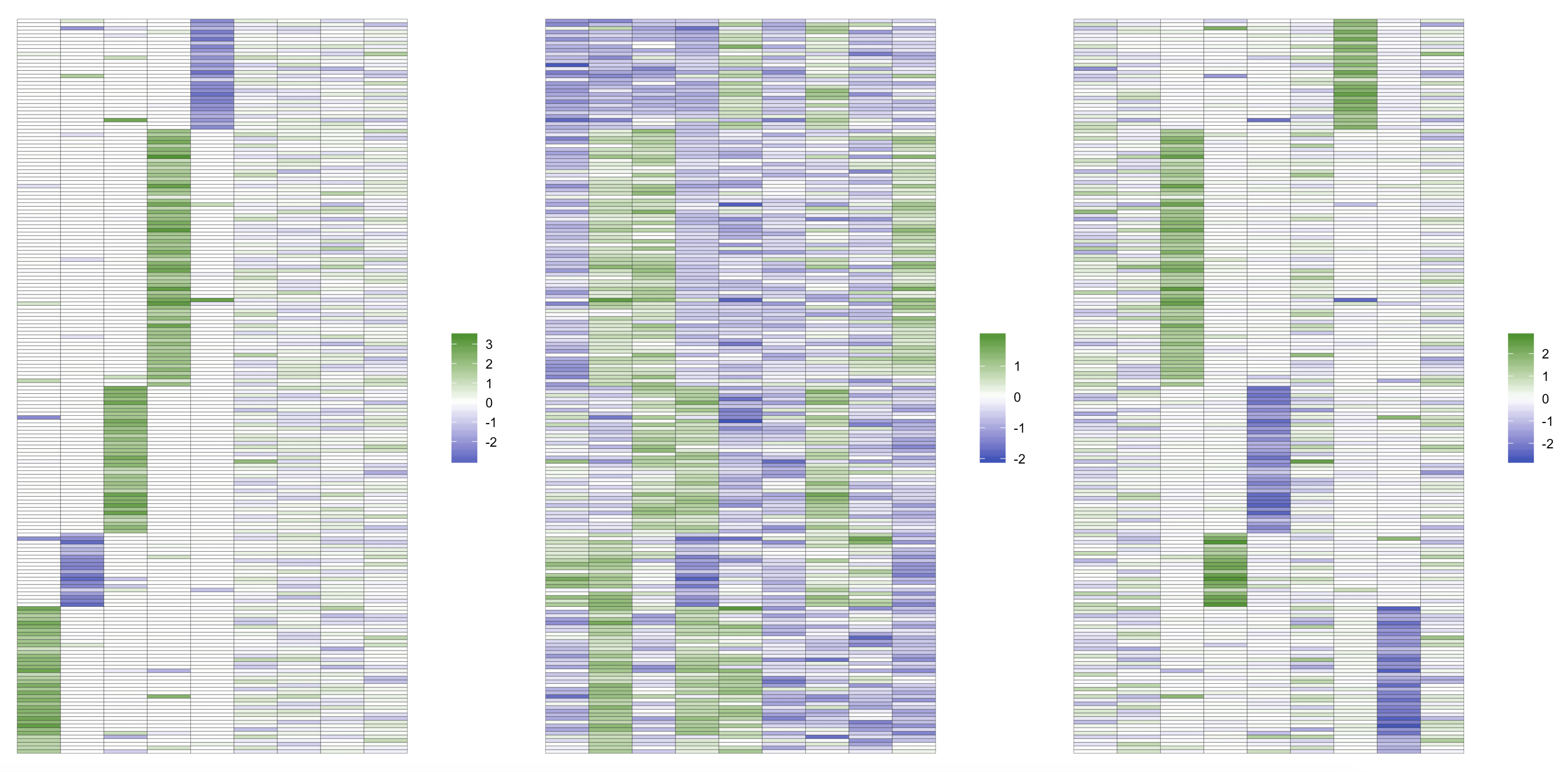}
    \caption{(Left) true factor loadings matrix $\Lambda$ used in simulation, (Center) a posterior sample of $\Lambda$ and (Right) the same sample of $\Lambda$ after applying orthogonalization. }
    \label{fig:True and Sampled Lambda}
\end{figure}

%The orthogonalization step is necessary to remove the sharing of related information across columns throughout the sampling process and has the benefit of representing the samples in a more interpretable form. In the factor analysis case, orthogonalization of significant factor loadings helps to cluster the covariates, often inducing some sparsity. Oblique rotations \citep{MFA} offer better representations of correlated factors, or more generally linear association in the columns of the unidentifiable matrix. However, the rotated matrices are not necessarily comparable from sample to sample. 
Let $\{\Lambda^{(t)}, \ t = 1,\ldots, T \}$ be the posterior samples of the factor loading matrix $\Lambda$. Our first goal is to tackle generic rotational invariance across $\Lambda^{(t)}$. In order to achieve that, we apply an orthogonalization procedure to each sample. Orthogonalization of factor loadings highlights the grouping of covariates, often inducing sparsity row-wise in $\Lambda$, which allows the samples to be represented in an interpretable form. Figure \ref{fig:True and Sampled Lambda} shows an example of the application of Varimax to a posterior sample of $\Lambda$. We define as $\Tilde{\Lambda}$ the factor loading matrix after applying Varimax. In our approach we focus on orthogonal rotations, and apply Varimax \citep{varimax} to each $\Lambda^{(t)}$. Oblique rotations \citep{MFA} can also provide accurate representations of correlated factors, or more generally linear association in the columns of the unidentifiable matrix. However, the obliquely rotated matrices would not necessarily be comparable across samples. %In fact, it is important that the rotational method does not normalize the samples by rows or columns. [{\bf DD - the text in the above part is again somewhat confusing - try to be more clear}]

%However, normalized representations can be valuable for exploring the structure of $\Lambda$ after applying orthogonalization. 

%\subsection{Choosing a Pivot} \label{sec::pivot}

After the orthogonalization step, the samples of $\Lambda$ still exhibit symptoms of non identifiability due to sign and permutation ambiguity in the columns \citep{conti2014bayesian}. Following \cite{papastamoulis2020identifiability}, we define $Q$ as a $k \times k$ permutation matrix. Each row and column of $Q$ has a single non-zero element, which is equal to $1$. We also define $S = \text{diag}(s_1,\ldots, s_k)$, with $s_j \in \{-1,1\}$, and $P = S Q$, a signed permutation matrix. Our goal is to find the signed permutation matrices in order to align the samples, and to do that we minimize the following loss function:
\begin{align}\label{loss_min}
    \text{minimize}_{Q^{(t)}, S^{(t)}; \ t = 1,\ldots, T} \sum_{t = 1}^T || \Tilde{\Lambda}^{(t)} Q^{(t)} S^{(t)} - \Lambda^P ||_F
\end{align}
where $||\cdot ||_F$ is the Frobenious norm, $\Lambda^P = [c_1^P \cdots \ c_k^P]$ is an exemplar factor loading matrix, which we call the pivot, and $c_j^{(t)}$ is the $j^{th}$ column of $\Tilde{\Lambda}^{(t)}$. We use the pivot to align the posterior samples, as in \cite{PRA}. The pivot matrix is used in the algorithm as a proxy of a `true' $\Lambda$, and we will provide the details of its choice later in this section. The loss function penalizes differences between the pivot and the posterior samples, and to minimize it we must find the signed rearrangement of $c_1^{(t)},\ldots,c_k^{(t)}$ that best matches the pivot's columns. %To do so, we match each $c_j^{(t)}$, for $j = 1, \ldots k$, with either $c_h^P$ or $-c_h^P$ for some $h = 1, \ldots k$. 

One possibility to minimize \eqref{loss_min} is to compute the loss for all possible signed permutations, which is computationally infeasible. Instead, we implement a greedy procedure and minimize the loss iteratively, i.e. column by column, as in \cite{PRA}.
%This creates the possibility that a column we reach earlier in our iteration incorrectly matches to a column that was closer in the 2-norm to a later column in the iteration. In such a case our algorithm would fail to find the true inverse permutation where PRA would but this is unlikely for columns that contain unique information. %{This creates ordering issues, so at each iteration we start matching from a randomly selected column.} This greedy search reduces the order of complexity from $\mathcal{O}(k!)$ to $\mathcal{O}(k\log k)$, a very significant decrease at minimal cost. \end{comment}
%We match the columns of the pivot to only column of $\Lambda^{(s)}$. This prevents to have multiple copies of a column of the pivot, which we call \textit{duplication}, in the aligned sample and no presence of other column.
%We use the vector $L_2$ norm to compare the columns of $\Lambda^{(s)}$ and $\Lambda^P$, which is the same norm used in the Varimax algorithm.
We compute the $L_2$ normed differences between the columns of $\Tilde{\Lambda}^{(t)}$ and the ones of $\Lambda^P$ and $-\Lambda^P$. We use the $L_2$ norm to match the objective of the Varimax algorithm. Other rotation schemes such as quartimax \citep{quartimax} may work better with the other norms. Our empirical simulations show that MatchAlign is not sensitive to the choice of norm. The computations are implemented in a greedy fashion: we compute the normed differences between the column of $\Tilde{\Lambda}^{(t)}$ with largest norm and $c_1^P, -c_1^P, \ldots, c_k^P, \ -c_k^P$. After matching with $c_j^P$ or $-c_j^P$, for some $j = 1, \ldots, k$ that minimizes the $L_2$ norm, we match the next column of $\Tilde{\Lambda}^{(t)}$ with $c_1^P, \ldots, c_{j-1}^P, -c_{j-1}^P,c_{j+1}^P, \ldots, -c_{k}^P$ and proceed iteratively. In this way, we only need to compute $k(2k+1)$ normed differences for each posterior sample, and the complexity of the matching step is $\mathcal{O}\big(T p k (2k + 1) \big)$. Whenever using a prior for $\Lambda$ that introduces increasing shrinkage as the column order increases, as in \cite{Bhattacharya2013} or \cite{legramanti2019bayesian}, we can naturally start matching the columns from $c_1^{(t)}$ and proceed in column order. 
%In practice we should only consider other orders of matching only if the prior choice on $\Lambda$ introduces decreasing shrinkage as the column order increases.

Due to sampling noise one $c_j^{(t)}$ may be minimally distant from more than one $c_h^P$. Because of that, we need to avoid matching multiple $c_j^{(t)}$s to one $c_h^P$, which we refer to as duplication. Alignment that involves duplicating some columns over others destroys information and prevents the accurate reconstruction of identifiable parameters such as the covariance after post-processing. Allowing duplication of columns would also introduce numerical instability when performing operations for posterior samples with duplicate columns, as well as biasing ergodic summaries of the parameters.

Several methods for factor models take an ``over-fitted'' approach \citep{Bhattacharya2013}, \citep{rousseau2011asymptotic}, and choose $k$ to correspond to an upper bound on the number of factors. In this scenario, we might have multiple columns centered near $0_p$. For this reason, our algorithm may not detect when these orthogonalized columns switch labels or signs between samples. However, this does not significantly bias the ergodic summaries of the factor loadings. In fact, columns might be mislabelled only if their normed difference is small. This is an important consideration because the matching method does not globally minimize differences between the reference and sample matrices. Instead, it operates on each column iteratively. 

In our algorithm, every column of $\Lambda^{(t)}$ is matched to either $c_j^P$ or $-c_j^P$, for some $j = 1,\ldots,k$. In order to correctly and efficiently match columns, we must choose a pivot that is central in the distribution of a column statistic.
%\textit{ A more central pivot will match more often with the members of its own distribution}. Estimates of the centers of the column distributions are then preferred.A single randomly sampled step in the MCMC is a consistent estimator for most measures of center. There is of course comparatively high variance in a single sample estimator but in practice this choice produces sensible results. If necessary a small number of samples can be manually aligned by sight and their sample mean or median can be used as a more informative pivot
As a proxy of unique information contained in each column, we consider the condition number:
\begin{align*}
    \kappa^{(t)} = \kappa(\Lambda^{(t)}) = \frac{\sigma_{max}(\Lambda^{(t)})}{\sigma_{min}(\Lambda^{(t)})}
\end{align*}
where $\sigma_{max}(\Lambda^{(t)})$ and $\sigma_{min}(\Lambda^{(t)})$ are the largest and smallest singular values of $\Lambda^{(t)}$, respectively. %An example of the distribution of $\kappa^{(1)}, \ldots, \kappa^{(T)}$ can be seen in Figure \ref{fig:condition_number}. 
We choose as a pivot the matrix with the median condition number. Note that this number may approach infinity with an overspecified number of columns. In this case, we can use the largest singular value in place of the condition number. We find in practice that these two choices provide similar performance of MatchAlign. One possibility to run the algorithm without selecting a pivot is by employing the solution of \cite{papastamoulis2020identifiability} based on the work of \cite{stephens2000dealing} in the context of mixture models. However, this solution would involve a significant increase in computational cost. Finally, notice that the identifiable parameters, such as $\Lambda \Lambda^T$, are not affected by applying MatchAlign, since we process the samples of $\Lambda$ by post-multiplying with a semi-orthogonal matrix.

\section{Simulations} \label{sec::simulation}

%In this section we compare the performance of the MatchAlign algorithm with the Varimax-RSP algorithm of \cite{papastamoulis2020identifiability}. In particular, we run their faster implementation with full simulated Annealing (full-SA). We generate the data according to \ref{fact_model} in two ways: 1) \textit{Independent}, by sampling each element of $\Lambda$ independently from a standard normal distribution and 2) \textit{Structured}, by dividing the covariates in three groups such that each of the groups mainly loads on a single latent factor.%, with the loading sampled from a Normal distribution with mean $1$ and variance $0.1$ and we sample the $k-3$ remaining columns of $\Lambda$ from a Normal distribution with mean $0$ and variance $0.1$. 

In this section we compare the performance of MatchAlign with three strategies of the Varimax-RSP algorithm of \cite{papastamoulis2020identifiability}: the Exact scheme (rsp exact), the fastest implementation with full simulated Annealing (RSP full-SA), and a hybrid implementation of the two strategies (RSP partial-SA). We generate $500$ data points according to \eqref{fact_model} in two ways: 1) \textit{Independent}, by sampling each element of $\Lambda$ independently from a standard normal distribution and 2) \textit{Sparse}, by dividing the covariates in three groups such that each of the groups mainly loads on a single latent factor. We generate the diagonal elements of $\Sigma$ from independent inverse-gamma distributions with parameters $(1/2, 1/2)$.

We fit a Gaussian factor model with a Dirichlet-Laplace prior \citep{DL} for each row of $\Lambda$ and an inverse-gamma prior with parameters $(1/2, 1/2)$ on the diagonal elements of the matrix $\Sigma$. We do not enforce any identifiability constraints on the matrix $\Lambda$. We use the function \texttt{linearDL} contained in \texttt{infinitefactor} \textbf{\textsf{R}} package to draw MCMC samples. We run the MCMC algorithm for $11000$ iterations and a burn-in of $1000$, keeping $10000$ in total. Notice that the run-times of the algorithms are independent of the number of observations since the input of the algorithms is $\{\Lambda^{(t)}: t=1,...,T\}$, which does not depend on $n$. We run the simulations for different values of $p$ and $k$, and for each set of parameters we average the run-times of the four algorithms over $25$ simulations. The alignment times were comparable over the \textit{Independent} and \textit{Sparse} settings so for this simulation we also average the results over these two scenarios.

\begin{figure}
    \centering
        \includegraphics[width=15cm, height=12cm]{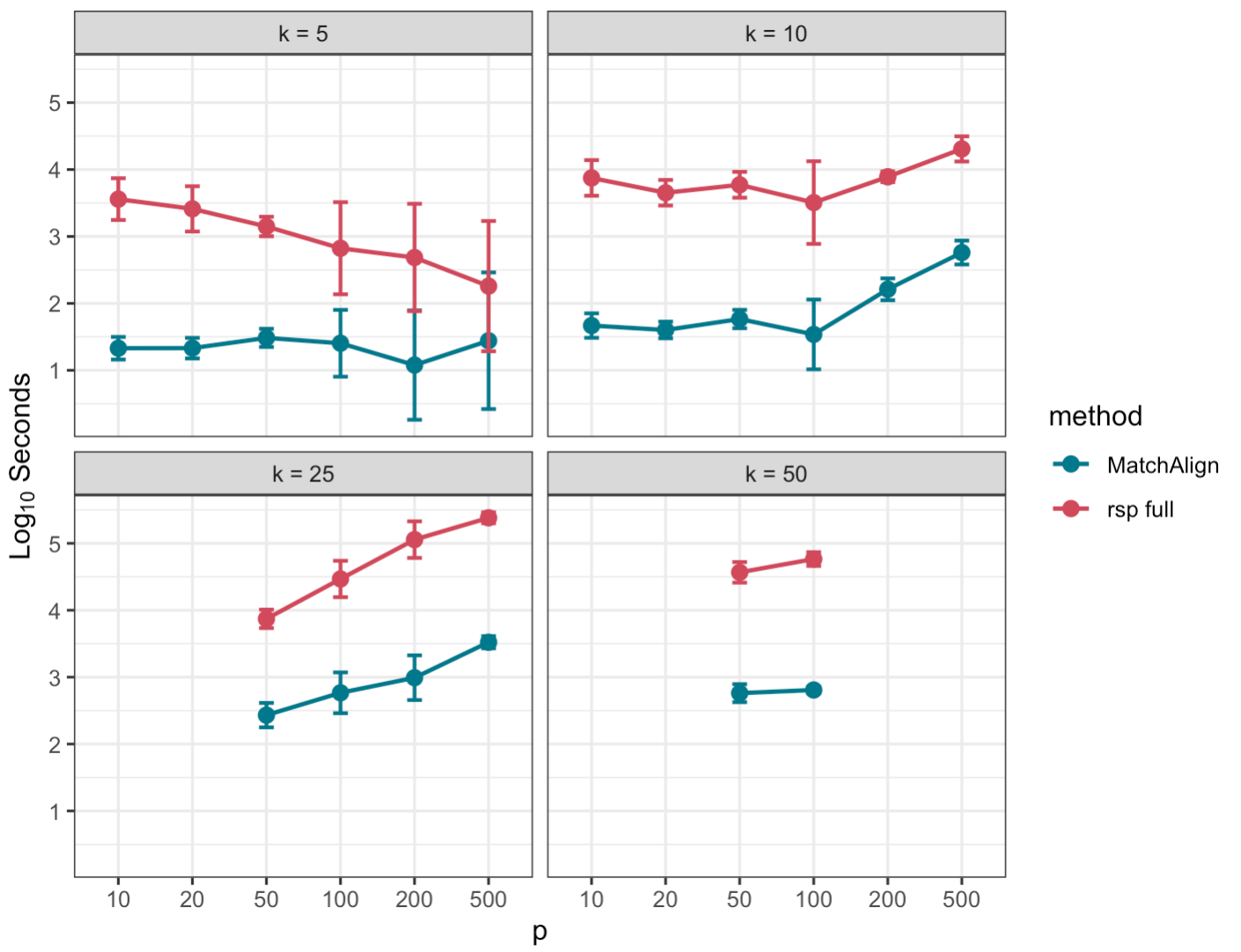}
    \caption{Comparison of running times in log$_{10}$ scale between MatchAlign and RSP with full-SA.}
    \label{MatchAlign_RSP}
\end{figure}

\begin{figure}
    \centering
        \includegraphics[width=15cm, height=12cm]{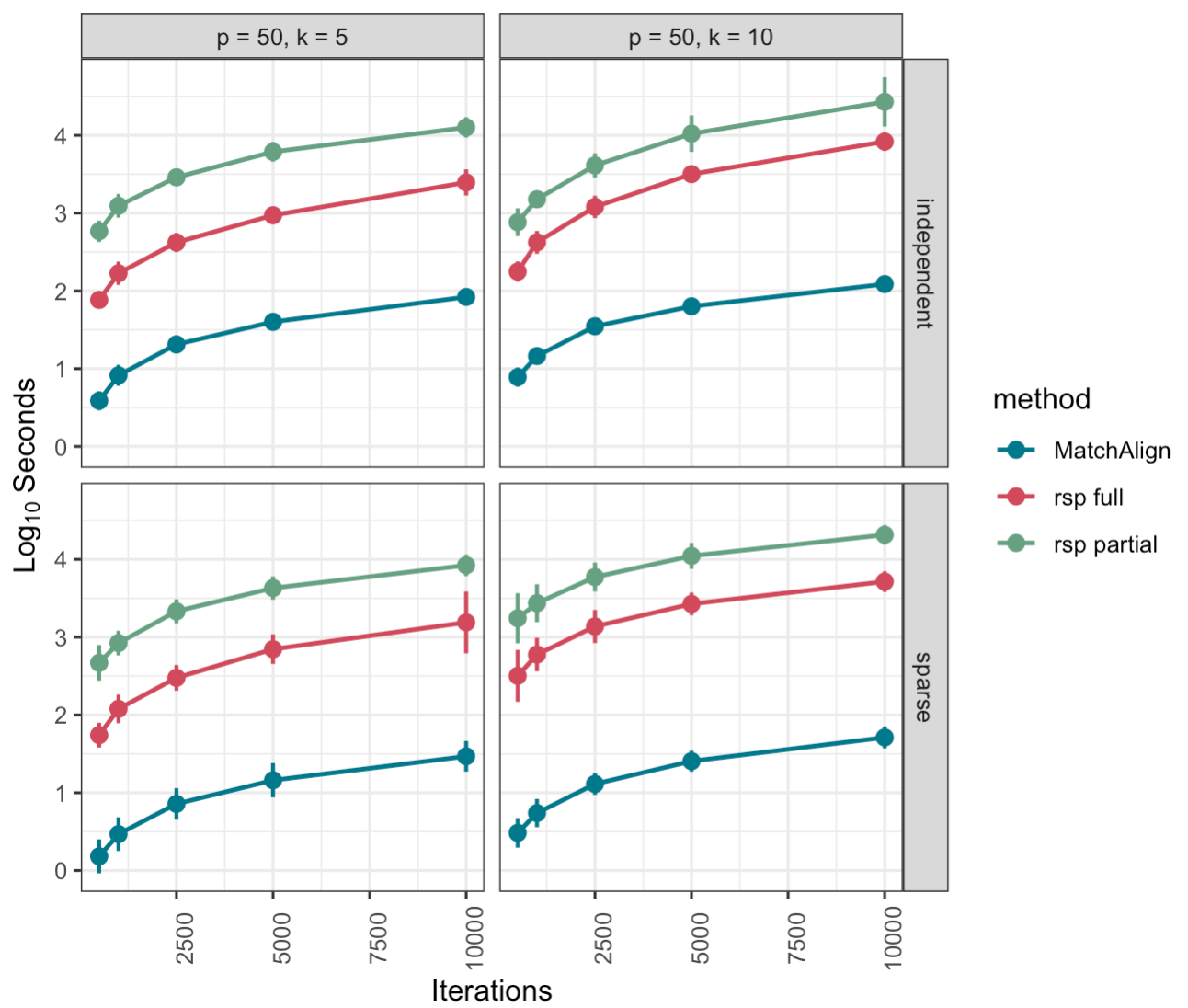}
    \caption{Comparison of running times in log$_{10}$ scale between MatchAlign and RSP with full-SA across multiple specifications of the number of MCMC iterations.}
    \label{MatchAlign_RSP_Iter}
\end{figure}

The results are shown in Figure \ref{MatchAlign_RSP} with the run-times on the $\text{log}_{10}$ scale. Notice how the MatchAlign algorithm is often at least an order of magnitude faster (approximately 10 times faster) than RSP full-SA when $k = 5,10$, and two orders of magnitude faster (approximately 100 times faster) when $k = 25,50$. In Figure \ref{MatchAlign_RSP_Iter} we show the run-times in log$_{10}$ scale between MatchAlign and RSP full-SA and RSP partial-SA across multiple specifications of the number of iterations taken by the MCMC sampler. Again, the difference between MatchAlign and the two implementations of the RSP algorithm is at least an order of magnitude.

%After comparing the running times of the algorithms, 
%We also propose a numeric assessment of the alignment goodness.
In order to evaluate the performance of the alignment procedures, one possibility is to use the normed difference between the posterior mean of the covariance and the estimated covariance using the posterior mean of the aligned $\Lambda$:
\begin{align}\label{metric_alignment}
    || \overline{\Lambda\Lambda^T} - \bar \Lambda_* \bar \Lambda_* ^T ||_F
\end{align}
where $\overline{\Lambda\Lambda^T} = \frac{1}{T}\sum_{t = 1}^T \Lambda^{(t)}(\Lambda^{(t)})^T$ and $\bar \Lambda_* = \frac{1}{T} \sum_{t = 1}^T \Lambda^{(t)}_*$ is the posterior mean of the aligned $\Lambda$. This follows from the intuition that inference under the rotated $\Lambda$ for the covariance matrix should not largely deviate from the posterior mean of the covariance, if the columns have been aligned correctly. The metric \eqref{metric_alignment} does not necessitate knowing the underlying true parameters, so it can be computed on real data as well. 

We compute metric \eqref{metric_alignment} on the simulated data for $k = 5,10$ and $p = 50,100$. We show the results in Table \ref{p_50_accuracy} and \ref{p_100_accuracy}. As expected, RSP Exact attains the best performance. However, MatchAlign's results are better than the fastest implementation of RSP and generally comparable to those of RSP Exact, which can only be run for $k \le 10$. 

We also compare the performance of the algorithms by estimating the effective sample size (ESS) after performing alignment. We average ESS over each element in $\Lambda$ and divide by the total number of samples. Generally, if the samples are not aligned correctly then the MCMC will experience poor mixing. We use ESS to compare the different methods but the ratio of ESS over the number of samples should not be close to one as, even with perfect alignment, MCMC algorithms will not produce uncorrelated samples from the posterior.
We show the results in Table \ref{p_50_ESS} and \ref{p_100_ESS}. Notice how MatchAlign performs similarly to the other algorithms while being at least one order of magnitude faster.

\begin{table}[!htbp] \centering 
  \caption{Comparison of MatchAlign, RSP full-SA, RSP partial-SA and RSP Exact when $p = 50$ according to metric \eqref{metric_alignment}. The metric values are presented as ratios compared to the best performing algorithm.} 
  \label{p_50_accuracy} 
\begin{tabular}{@{\extracolsep{5pt}} c|c|cccc} 
\\[-1.8ex]\hline 
\hline 
& & MatchAlign & RSP exact & RSP full & RSP partial \\ 
\hline
$\Lambda$ sparse & k=5 & $3.548$ & $1$ & $3.563$ & $1.000$ \\ 
& k=10 & $1.331$ & $1$ & $3.941$ & $1.134$ \\ 
\hline 
$\Lambda$ independent & k=5 & $1$ & $1.033$ & $1.613$ & $1.033$ \\ 
& k=10 & $1.176$ & $1$ & $1.488$ & $1.009$ \\ 
\hline \\[-1.8ex] 
\end{tabular} 
\end{table}

\begin{table}[!htbp] \centering 
  \caption{Comparison of MatchAlign, RSP full-SA, RSP partial-SA and RSP Exact when $p = 100$ according to metric \eqref{metric_alignment}. The metric values are presented as ratios compared to the best performing algorithm.} 
  \label{p_100_accuracy} 
\begin{tabular}{@{\extracolsep{5pt}} c|c|cccc} 
\\[-1.8ex]\hline 
\hline 
& & MatchAlign & RSP exact & RSP full & RSP partial \\ 
\hline 
$\Lambda$ sparse & k=5 & $2.835$ & $1$ & $1.191$ & $1.000$ \\ 
& k=10 & $1.303$ & $1$ & $2.501$ & $1.053$ \\ 
\hline 
$\Lambda$ independent &k=5 & $1.033$ & $1$ & $2.262$ & $1.001$ \\ 
& k=10 & $1.260$ & $1$ & $1.321$ & $1.010$ \\ 
\hline \\[-1.8ex] 
\end{tabular} 
\end{table}

\begin{table}[!htbp] \centering 
  \caption{Comparison of the MatchAlign, RSP full-SA, RSP partial-SA and RSP Exact when $p = 50$ using ESS divided by the total number of samples.} 
  \label{p_50_ESS} 
\begin{tabular}{@{\extracolsep{5pt}} c|c|cccc} 
\\[-1.8ex]\hline 
\hline 
& & MatchAlign & RSP exact & RSP full & RSP partial \\ 
\hline 
$\Lambda$ sparse & k = 5 & $0.686$ & $0.630$ & $0.646$ & $0.635$ \\ 
& k = 10 & $0.797$ & $0.738$ & $0.826$ & $0.783$ \\ 
\hline
$\Lambda$ independent & k = 5 & $0.465$ & $0.522$ & $0.496$ & $0.525$ \\ 
& k = 10 & $0.380$ & $0.417$ & $0.523$ & $0.423$ \\ 
\hline \\[-1.8ex] 
\end{tabular} 
\end{table}

\begin{table}[!htbp] \centering 
  \caption{Comparison of MatchAlign, RSP full-SA, RSP partial-SA and RSP Exact when $p = 100$ using ESS divided by the total number of samples.} 
  \label{p_100_ESS} 
\begin{tabular}{@{\extracolsep{5pt}} c|c|cccc} 
\\[-1.8ex]\hline 
\hline 
 & & MatchAlign & rsp exact & rsp full & rsp partial \\ 
\hline \\[-1.8ex] 
$\Lambda$ sparse & k = 5 & $0.728$ & $0.641$ & $0.644$ & $0.641$ \\ 
& k = 10 & $0.819$ & $0.782$ & $0.854$ & $0.789$ \\ 
\hline
$\Lambda$ independent & k = 5 & $0.312$ & $0.305$ & $0.325$ & $0.307$ \\ 
& k = 10 & $0.257$ & $0.264$ & $0.357$ & $0.276$ \\ 
\hline \\[-1.8ex] 
\end{tabular} 
\end{table}

\section{Application} \label{sec::genomic}

We are particularly motivated by studies of environmental health collecting data on mixtures of chemical exposures. These exposures can be moderately high-dimensional with high correlations within blocks of variables; for example, this can arise when an individual is exposed to a product having a mixture of chemicals and when chemical measurements consist of metabolites or breakdown products of a parent compound. In this application, we use data from the National Health and Nutrition Examination Survey (NHANES) collected in 2015 and 2016. We select a subset of $107$ chemical exposures for which at least $30\%$ of measurements have been recorded (i.e., are not missing). We also select a subsample of $4468$ individuals for which at least $30\%$ of the $107$ chemicals have been recorded. After this preprocessing step, we are left with a matrix $X$ of dimension $4468 \times 107$ with $36\%$ missing data.

We implement an initial Bayesian analysis of model \eqref{fact_model} with the prior of \cite{Bhattacharya2013} for $\Lambda$ and an inverse-gamma prior with parameters $(1/2, 1/2)$ on the diagonal elements of the matrix $\Sigma$, as in the simulation experiments. We notice from the eigen-decomposition of the correlation matrix that the first $25$ eigenvectors explain more than $80\%$ of the total variability; hence, we set the number of factors equal to $25$. We run the MCMC procedure for $11000$ iterations with a burn-in of $10 000$ and apply the MatchAlign and RSP-full algorithms to the posterior samples of $\Lambda$. 

\begin{figure}[h]
    \centering
    \includegraphics[width=15cm, height=16cm]{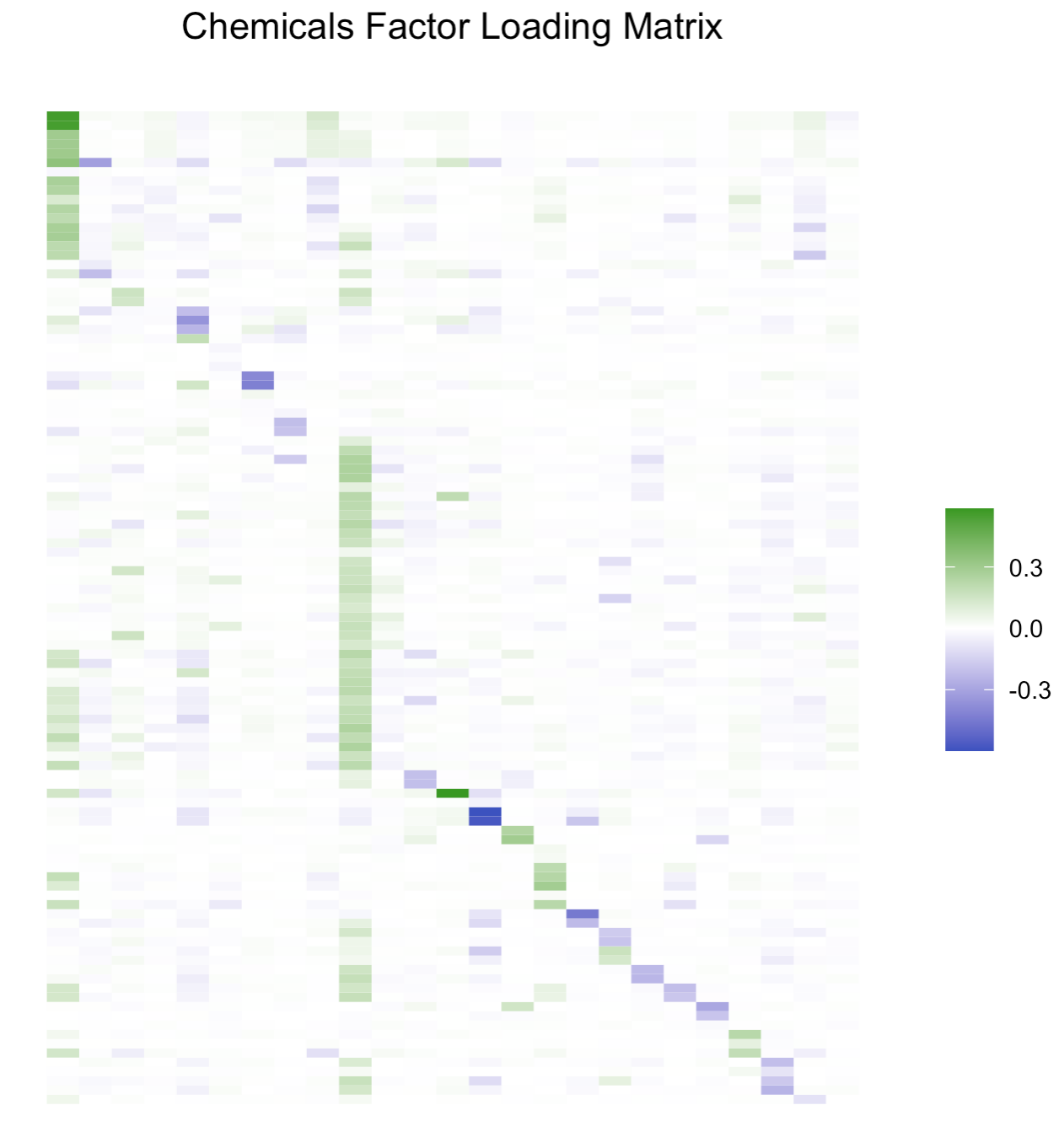}
    \caption{Matrix of factor loadings for NHANES 2015-2016 dataset after postprocessing with MatchAlign algorithm.}
    \label{lambda_application}
\end{figure}

\begin{figure}[h]
    \begin{minipage}{.5\textwidth}
    \vspace{.45cm}
        \centering
        \includegraphics[width=7cm, height=5.5cm]{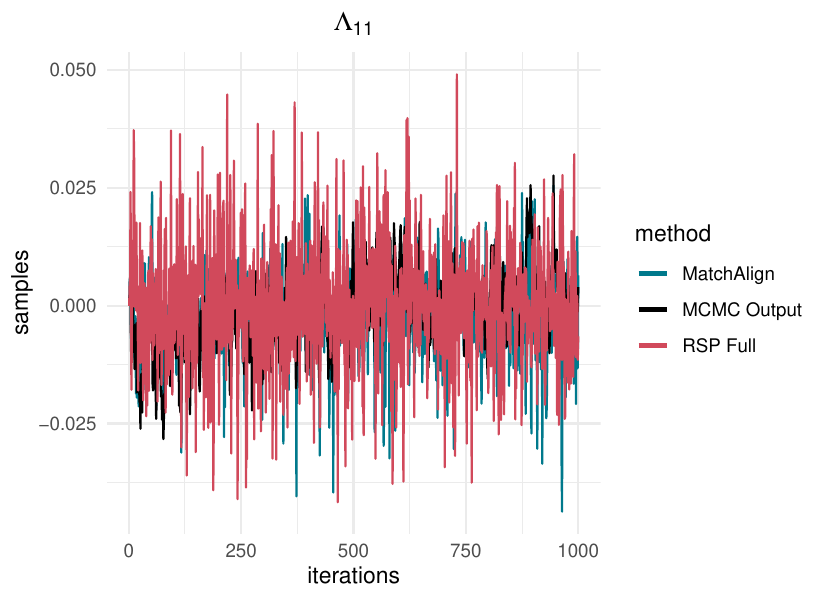}
    \end{minipage}%
    \begin{minipage}{.5\textwidth}
        \centering
        \includegraphics[width=7cm, height=5.5cm]{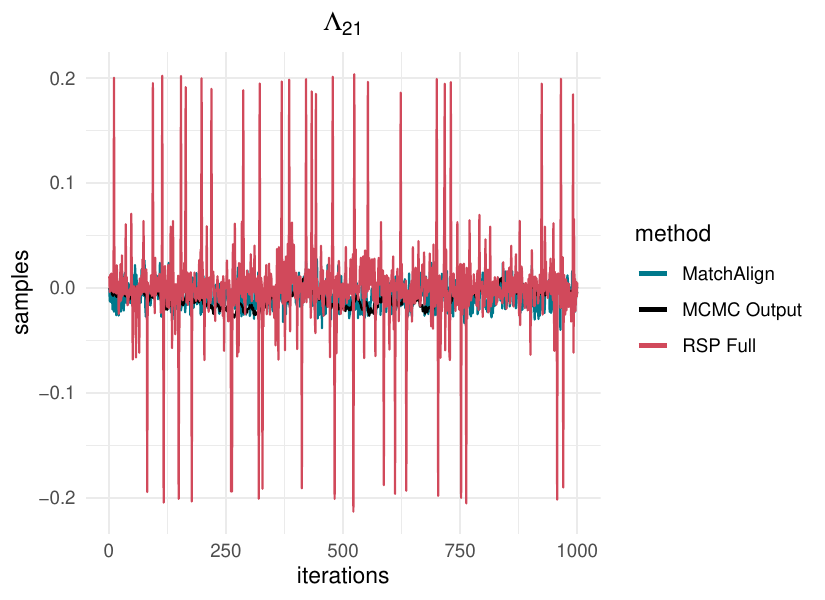}
    \end{minipage}%
    
    \hfill

    \begin{minipage}{.5\textwidth}
    \vspace{.45cm}
        \centering
        \includegraphics[width=7cm, height=5.5cm]{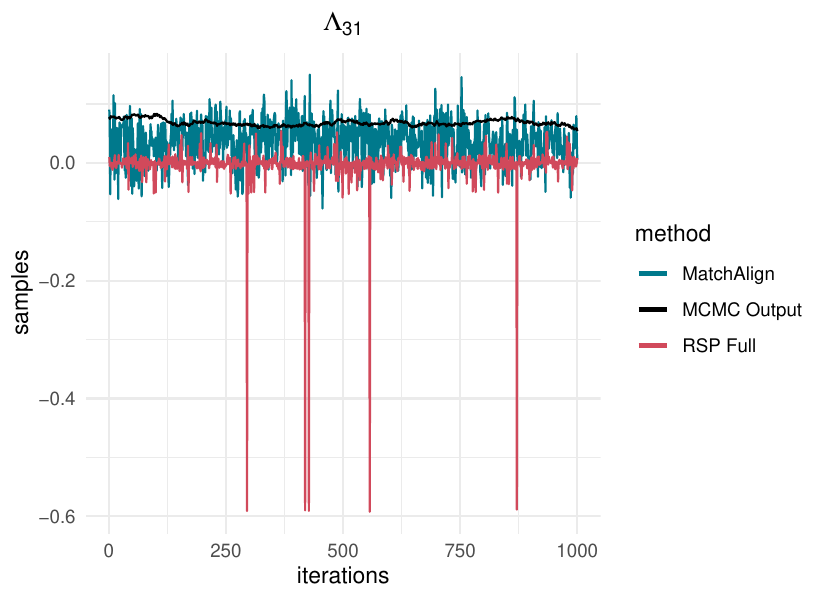}
    \end{minipage}%
    \begin{minipage}{.5\textwidth}
        \centering
        \includegraphics[width=7cm, height=5.5cm]{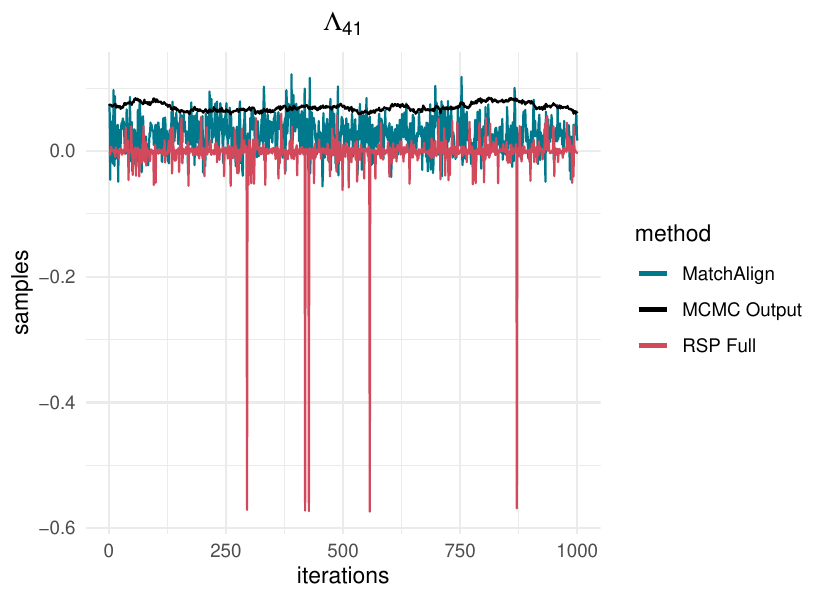}
    \end{minipage}%
    
    \hfill
    
    \begin{minipage}{.5\textwidth}
    \vspace{.45cm}
        \centering
        \includegraphics[width=7cm, height=5.5cm]{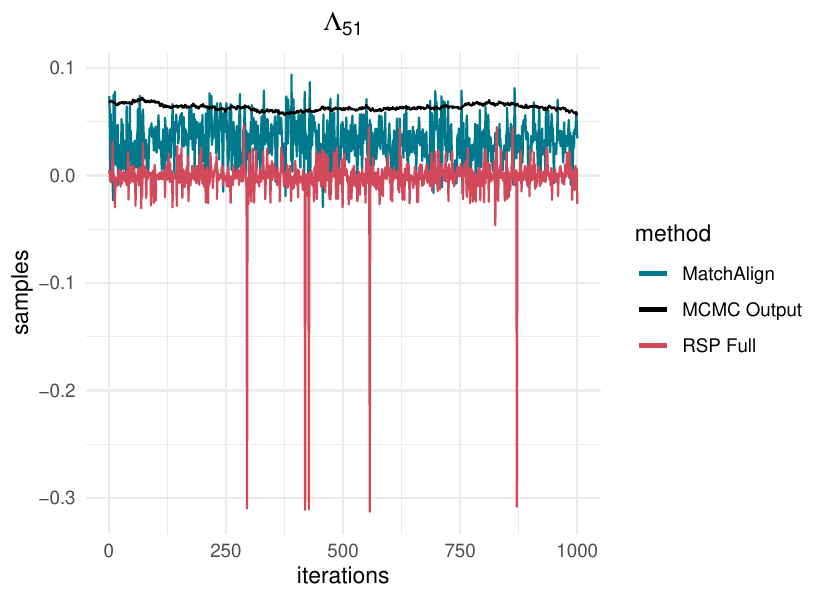}
    \end{minipage}%
    \begin{minipage}{.5\textwidth}
        \centering
        \includegraphics[width=7cm, height=5.5cm]{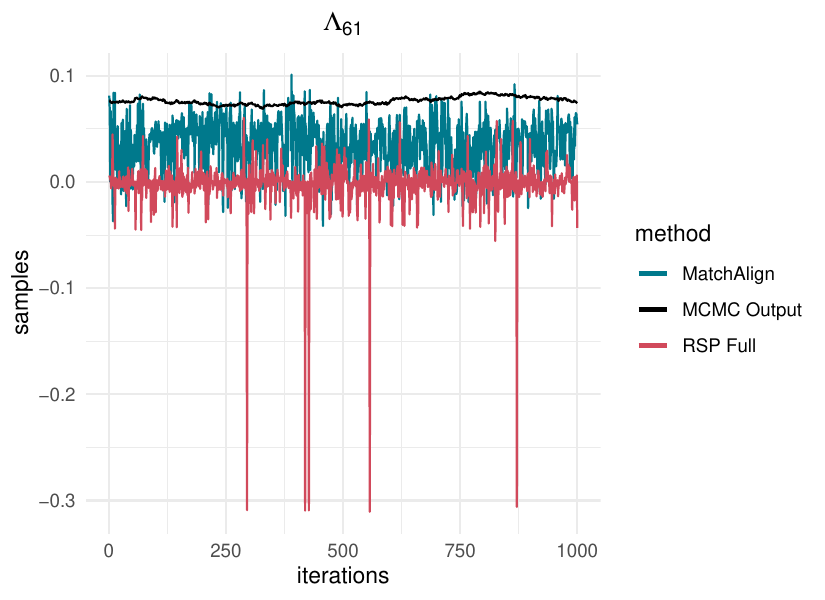}\\
    \end{minipage}%
    \caption{Traceplots of the factor loadings $\Lambda_{11},\ldots,\Lambda_{16}$ before and after rotation with MatchAlign and RSP-Full.}
    \label{Traceplots_application}
\end{figure}

Figure \ref{lambda_application} shows the aligned posterior mean for the factor loadings matrix for the $107$ chemical exposures. We re-ordered the rows of $\Lambda$ to highlight the chemical groupings. The time to align the samples using MatchAlign was around $16$ seconds on a single thread of a 4.5GHz processor. %The scaling with $p$ is limited to the increased computation of taking the 2-norm of the differences, but this can be efficiently parallelized using openBLAS or other multi-threaded linear algebra libraries. 
Figure \ref{Traceplots_application} shows the traceplots of $\Lambda_{11},\ldots,\Lambda_{16}$ before and after rotation. RSP-Full fails to properly align some samples resulting in biased estimates. The value of the metric \eqref{metric_alignment} was approximately $20$ times higher for RSP-Full compared to MatchAlign.

\section{Discussion}

We proposed a computationally efficient post-processing algorithm that solves label and sign switching in matrix valued parameters that are subject to rotational ambiguity. Solving rotational ambiguity for high-dimensional parameters in Bayesian models can be challenging because of the of high computational cost. Our MatchAlign algorithm reduces the solution space by iteratively comparing columns, saving massive computational time in comparison to permutation searches, as in \cite{papastamoulis2020identifiability}. In Section \ref{sec::simulation} we compared the computational performance of MatchAlign with respect to the algorithm of \cite{papastamoulis2020identifiability} and show that MatchAlign is at least one order of magnitude faster, without compromising the performance of alignment. %We also show that MatchAlign can be applied for high-dimensional applications in Section \ref{sec::genomic}.

Crucially, this algorithm does not alter the estimation of identifiable parameters, such as $\Lambda \Lambda^T$, and allows one to perform inference on the factor loadings matrix. The code is available in the \texttt{infinitefactor} \textbf{\textsf{R}} package on the CRAN repository and can be easily applied in broad settings.

%We have made steps in improving the efficiency in resolving label switching, but there are additional challenges that could be solved using MatchAlign. For example, this method is directly applicable to row-switching in matrix samples by taking a transpose before and after performing alignment, and method could be quickly generalized to symmetric matrices that display both row and column switching.

\section*{Acknowledgments}

This research was supported by grant 1R01ES028804-01 of the National Institute of Environmental Health Sciences of the United States Institutes of Health. F.F. was partially supported by Project Data Sphere and grant 1R01LM013352-01A1 of the National Institutes of Health. The authors would like to thank Kelly Moran, Melody Jiang and Bianca Dumitrascu for helpful comments.

\bibliographystyle{Chicago}
\bibliography{LS}

\begin{thebibliography}{}

\bibitem[\protect\citeauthoryear{Anderson, Rubin, et~al.}{Anderson
  et~al.}{1956}]{anderson1956statistical}
Anderson, T.~W., H.~Rubin, et~al. (1956).
\newblock Statistical inference in factor analysis.
\newblock In {\em Proceedings of the third Berkeley symposium on mathematical
  statistics and probability}, Volume~5, pp.\  111--150.

\bibitem[\protect\citeauthoryear{Aßmann, Boysen-Hogrefe, and Pape}{Aßmann
  et~al.}{2016}]{BADFM}
Aßmann, C., J.~Boysen-Hogrefe, and M.~Pape (2016).
\newblock Bayesian analysis of static and dynamic factor models: An ex-post
  approach towards the rotation problem.
\newblock {\em Journal of Econometrics\/}~{\em 192\/}(1), 190--206.

\bibitem[\protect\citeauthoryear{Bhattacharya and Dunson}{Bhattacharya and
  Dunson}{2011}]{Bhattacharya2013}
Bhattacharya, A. and D.~B. Dunson (2011).
\newblock {Sparse Bayesian infinite factor models}.
\newblock {\em Biometrika\/}~{\em 98}, 291--306.

\bibitem[\protect\citeauthoryear{Bhattacharya, Pati, Pillai, and
  Dunson}{Bhattacharya et~al.}{2015}]{DL}
Bhattacharya, A., D.~Pati, N.~S. Pillai, and D.~B. Dunson (2015).
\newblock Dirichlet–laplace priors for optimal shrinkage.
\newblock {\em Journal of the American Statistical Association\/}~{\em
  110\/}(512), 1479--1490.

\bibitem[\protect\citeauthoryear{Carvalho, Chang, Lucas, Nevins, Wang, and
  West}{Carvalho et~al.}{2008}]{carvalho2008high}
Carvalho, C.~M., J.~Chang, J.~E. Lucas, J.~R. Nevins, Q.~Wang, and M.~West
  (2008).
\newblock High-dimensional sparse factor modeling: applications in gene
  expression genomics.
\newblock {\em Journal of the American Statistical Association\/}~{\em
  103\/}(484), 1438--1456.

\bibitem[\protect\citeauthoryear{Chen, Li, and Zhang}{Chen
  et~al.}{2020}]{chen2020structured}
Chen, Y., X.~Li, and S.~Zhang (2020).
\newblock Structured latent factor analysis for large-scale data:
  Identifiability, estimability, and their implications.
\newblock {\em Journal of the American Statistical Association\/}~{\em
  115\/}(532), 1756--1770.

\bibitem[\protect\citeauthoryear{Conti, Fr{\"u}hwirth-Schnatter, Heckman, and
  Piatek}{Conti et~al.}{2014}]{conti2014bayesian}
Conti, G., S.~Fr{\"u}hwirth-Schnatter, J.~J. Heckman, and R.~Piatek (2014).
\newblock Bayesian exploratory factor analysis.
\newblock {\em Journal of econometrics\/}~{\em 183\/}(1), 31--57.

\bibitem[\protect\citeauthoryear{Erosheva and Curtis}{Erosheva and
  Curtis}{2011}]{erosheva2011dealing}
Erosheva, E.~A. and S.~M. Curtis (2011).
\newblock Dealing with rotational invariance in bayesian confirmatory factor
  analysis.
\newblock {\em Department of Statistics, University of Washington, Seattle,
  Washington, USA\/}.

\bibitem[\protect\citeauthoryear{Fr{\"u}hwirth-Schnatter and
  Lopes}{Fr{\"u}hwirth-Schnatter and Lopes}{2018}]{fruhwirth2018sparse}
Fr{\"u}hwirth-Schnatter, S. and H.~F. Lopes (2018).
\newblock Sparse bayesian factor analysis when the number of factors is
  unknown.
\newblock {\em arXiv preprint arXiv:1804.04231\/}.

\bibitem[\protect\citeauthoryear{Geweke and Zhou}{Geweke and
  Zhou}{1996}]{geweke1996measuring}
Geweke, J. and G.~Zhou (1996).
\newblock Measuring the pricing error of the arbitrage pricing theory.
\newblock {\em The Review of Financial Studies\/}~{\em 9\/}(2), 557--587.

\bibitem[\protect\citeauthoryear{Ghosh and Dunson}{Ghosh and
  Dunson}{2009}]{Ghosh}
Ghosh, J. and D.~B. Dunson (2009).
\newblock Default prior distributions and efficient posterior computation in
  bayesian factor analysis.
\newblock {\em Journal of Computational and Graphical Statistics\/}~{\em
  18\/}(2), 306--320.
\newblock PMID: 23997568.

\bibitem[\protect\citeauthoryear{Kaiser}{Kaiser}{1958}]{varimax}
Kaiser, H.~F. (1958, Sep).
\newblock The varimax criterion for analytic rotation in factor analysis.
\newblock {\em Psychometrika\/}~{\em 23\/}(3), 187--200.

\bibitem[\protect\citeauthoryear{Kaufmann and Schumacher}{Kaufmann and
  Schumacher}{2017}]{kaufmann2017identifying}
Kaufmann, S. and C.~Schumacher (2017).
\newblock Identifying relevant and irrelevant variables in sparse factor
  models.
\newblock {\em Journal of Applied Econometrics\/}~{\em 32\/}(6), 1123--1144.

\bibitem[\protect\citeauthoryear{Kaufmann and Schumacher}{Kaufmann and
  Schumacher}{2019}]{kaufmann2019bayesian}
Kaufmann, S. and C.~Schumacher (2019).
\newblock Bayesian estimation of sparse dynamic factor models with
  order-independent and ex-post mode identification.
\newblock {\em Journal of Econometrics\/}~{\em 210\/}(1), 116--134.

\bibitem[\protect\citeauthoryear{Lawley and Maxwell}{Lawley and
  Maxwell}{1962}]{lawley1962factor}
Lawley, D.~N. and A.~E. Maxwell (1962).
\newblock Factor analysis as a statistical method.
\newblock {\em Journal of the Royal Statistical Society: Series D (The
  Statistician)\/}~{\em 12\/}(3), 209--229.

\bibitem[\protect\citeauthoryear{Legramanti, Durante, and Dunson}{Legramanti
  et~al.}{2020}]{legramanti2019bayesian}
Legramanti, S., D.~Durante, and D.~B. Dunson (2020).
\newblock Bayesian cumulative shrinkage for infinite factorizations.
\newblock {\em Biometrika\/}~{\em 107\/}(3), 745--752.

\bibitem[\protect\citeauthoryear{Lorenzo-Seva}{Lorenzo-Seva}{1999}]{promin}
Lorenzo-Seva, U. (1999).
\newblock Promin: A method for oblique factor rotation.
\newblock {\em Multivariate Behavioral Research\/}~{\em 34\/}(3), 347--365.

\bibitem[\protect\citeauthoryear{Lucas, Carvalho, Wang, Bild, Nevins, and
  West}{Lucas et~al.}{2006}]{lucas2006sparse}
Lucas, J., C.~Carvalho, Q.~Wang, A.~Bild, J.~R. Nevins, and M.~West (2006).
\newblock Sparse statistical modelling in gene expression genomics.
\newblock {\em Bayesian Inference for Gene Expression and Proteomics\/}~{\em
  1\/}(1).

\bibitem[\protect\citeauthoryear{Marin, L.~Mengersen, and Robert}{Marin
  et~al.}{2005}]{PRA}
Marin, J.-M., K.~L.~Mengersen, and C.~Robert (2005, 12).
\newblock Bayesian modelling and inference on mixtures of distributions.
\newblock {\em Handbook of Statistics\/}~{\em 25}.

\bibitem[\protect\citeauthoryear{McAlinn, Rockova, and Saha}{McAlinn
  et~al.}{2018}]{mcalinn2018dynamic}
McAlinn, K., V.~Rockova, and E.~Saha (2018).
\newblock Dynamic sparse factor analysis.
\newblock {\em arXiv preprint arXiv:1812.04187\/}.

\bibitem[\protect\citeauthoryear{Millsap}{Millsap}{2001}]{millsap2001trivial}
Millsap, R.~E. (2001).
\newblock When trivial constraints are not trivial: The choice of uniqueness
  constraints in confirmatory factor analysis.
\newblock {\em Structural Equation Modeling\/}~{\em 8\/}(1), 1--17.

\bibitem[\protect\citeauthoryear{Neuhaus and Wrigley}{Neuhaus and
  Wrigley}{1954}]{quartimax}
Neuhaus, J.~O. and C.~Wrigley (1954).
\newblock The quartimax method: an analytic approach to orthogonal simple
  structure.
\newblock {\em British Journal of Statistical Psychology\/}~{\em 7\/}(2),
  81--91.

\bibitem[\protect\citeauthoryear{Papastamoulis and Ntzoufras}{Papastamoulis and
  Ntzoufras}{2020}]{papastamoulis2020identifiability}
Papastamoulis, P. and I.~Ntzoufras (2020).
\newblock On the identifiability of {Bayesian} factor analytic models.
\newblock {\em arXiv preprint arXiv:2004.05105\/}.

\bibitem[\protect\citeauthoryear{Rousseau and Mengersen}{Rousseau and
  Mengersen}{2011}]{rousseau2011asymptotic}
Rousseau, J. and K.~Mengersen (2011).
\newblock Asymptotic behaviour of the posterior distribution in overfitted
  mixture models.
\newblock {\em Journal of the Royal Statistical Society: Series B (Statistical
  Methodology)\/}~{\em 73\/}(5), 689--710.

\bibitem[\protect\citeauthoryear{Stephens}{Stephens}{2000}]{stephens2000dealing}
Stephens, M. (2000).
\newblock Dealing with label switching in mixture models.
\newblock {\em Journal of the Royal Statistical Society: Series B (Statistical
  Methodology)\/}~{\em 62\/}(4), 795--809.

\bibitem[\protect\citeauthoryear{Thurstone}{Thurstone}{1931}]{MFA}
Thurstone, L.~L. (1931).
\newblock Multiple factor analysis.
\newblock {\em Psychological review\/}~{\em 38\/}(5), 406.

\end{thebibliography}

\end{document}